\documentclass{aa}
\usepackage{graphicx,epsfig}
\usepackage{amssymb,amsmath}
\usepackage{xspace}
\usepackage{natbib}
\bibpunct{(}{)}{;}{a}{}{,}

\newcommand{\Ha}{H$_\alpha$\xspace}
\newcommand{\mHa}{\ensuremath{\mathrm{H}_\alpha}}
\newcommand{\LHa}{$L(\mHa)$\xspace}

\newcommand{\cfac}{\ensuremath{{\cal V}^2}}
\newcommand{\tcfac}{$\cfac$\xspace}

\begin{document}

\title{On Star Formation Rates in Dwarf Galaxies}
\subtitle{}
\author{Peter~M.~Weilbacher, Uta~Fritze-v.Alvensleben}
\institute{Universit\"ats-Sternwarte, Geismarlandstr.~11, 37083 G\"ottingen, Germany}
\authorrunning{P.~Weilbacher \& U.~Fritze-v.Alvensleben}
\titlerunning{SFR in Dwarf Galaxies}
\offprints{P.~Weilbacher, \email{weilbach@uni-sw.gwdg.de}}
\date{Received 7 May 2001 / Accepted 16 May 2001}

\abstract{We present evolutionary synthesis models of starbursts on
  top of old stellar populations to investigate in detailed time
  evolution the relation between \Ha luminosity and star formation
  rate (SFR).  The models show that several effects have an impact on
  the ratio between \LHa and SFR. Metallicity different from solar
  abundance, a time delay between star formation and maximum
  \Ha-luminosity, and a varying stellar initial mass function give rise
  to strong variations in the ratio of \Ha luminosity to SFR and can
  cause large errors in the determination of the SFR when employing
  well-known calibrations.  When studying star-bursting dwarf
  galaxies, and sub-galactic fragments at high redshift, which show
  SFR fluctuating on short timescales, these effects can add up to
  errors of two orders of magnitude compared with the calibrations. To
  accurately determine the true current SFR additional data in
  combination with models for the spectral energy distribution are
  needed.
  \keywords{Galaxies: starbursts -- galaxies: dwarf -- galaxies:
    evolution -- galaxies: fundamental parameters}
}

\maketitle
%
%________________________________________________________________

\section{Introduction}\label{sec:intro}

The star formation rate (SFR) is one of the basic properties of
galaxies.  It can be derived in different wavelength regimes (UV,
optical, FIR) using empirical calibrations obtained from well studied
samples of galaxy types, like spirals, irregulars, starbursts, or
luminous IR galaxies.  One of the most popular methods to derive SFRs
from optical observations is by measuring the Balmer line fluxes,
which are very sensitive to the \ion{H}{ii} regions surrounding
massive young stars and therefore give a good measure of the very
recent, shortlived, or ongoing SFR. When employing \Ha one generally
takes one of the simple linear calibrations as e.g.~obtained by
\citet[][ hereafter HG86]{HG86} or \citet[][ KTC94]{KTC94} from
observations and modeling of dwarf irregular and spiral galaxies,
respectively. The calibrations are used in the form\footnote{The
  name of the variable $\cfac = L(\mHa) / $SFR is deduced from the
  fact, that the unit of \tcfac actually is a squared velocity.}
\begin{equation}
   L(\mHa) \, \left[\mathrm{erg\,s}^{-1}\right] = \cfac \cdot \mathrm{SFR}\,\left[\mathrm{M}_\odot\,\mathrm{yr}^{-1}\right],
   \label{eq:Ha}
\end{equation}
where the \tcfac-factor is $1.41\cdot10^{41}$ (HG86) or
$1.26\cdot10^{41}$ \citep[KTC94, ][]{Ken98b} for a Salpeter IMF in the
mass range of 0.1 to 100\,M$_\odot$.

%mostly quiet systems with a more or less constant low-level 
These \LHa-SFR calibrations were derived for ``normal'' galaxies with
modest star formation rates. For systems which behave differently from
the ones for which the calibrations were derived, the application of
this method may not be appropriate. The validity of the
SFR-determination in star-bursting galaxies, like e.g.~blue compact
dwarfs (BCDs), where these relations are frequently used, has never
been shown.

In fact, as the strongest output of ionizing photons is related to the
most luminous, i.e.~giant or supergiant phases of the ionizing stars,
there may be a small time delay between abrupt changes in the SFR and
the corresponding changes in the \Ha flux. \citet{LRH95} showed that
such a time dependence exists for stellar features in the UV. For
systems with SF fluctuations on short timescales, the delay effect of
the Balmer lines can also be important, e.g.~for small scale systems
like dwarf galaxies, where star burst durations are usually assumed to
be of the order of a dynamical timescale, i.e.~10$^5$ to 10$^6$ years.
SFR fluctuating strongly on short timescales may also have taken place
in sub-galactic fragments before merging together to hierarchically
build up today's galaxies \citep{GBE+99}.

We will show that indeed one may make very large errors when blindly
applying the usual calibrations to small systems. We first present
details on our model in Sect.~\ref{sec:model}. We then describe
different effects that can affect the calibrations for \LHa in terms
of SFR, namely metallicity (Sect.~\ref{sec:metal}), short burst durations
(Sect.~\ref{sec:shortb}), and changes in the IMF
(Sect.~\ref{sec:imf}). We finally summarize our results in
Sect.~\ref{sec:concl}.

%__________________________________________________________________

\vspace*{-5pt}
\section{Model description}\label{sec:model}

We use our evolutionary synthesis code specifically adapted to the
modeling of starbursts in dwarf galaxies \citep{KFL95,WDF+00}. It
includes specific modeling of gaseous emission lines and continuum
based on the Lyman continuum photons emitted by hot young stars.

%{\bf wir sollten auch die neuen Tracks mit $Z = 1/50 Z_\odot$ mitnehmen..., Probleme beim Einbau, vielleicht doch nicht...}
We use the current Geneva stellar tracks \citep[see][ for a recent
compilation]{LS01} in the metallicity range from $Z_\odot/$20 to
$2\,Z_\odot$.  To be compatible with HG86 and KTC94 we use the
Salpeter IMF \citep{Sal55} in the range of 0.15 to 85 or 120 M$_\odot$, as given by
the tracks. We use the recently revised Lyman continuum photon
emission rates as given by \citet{SdK97}, who account for non-LTE
effects, line blanketing, stellar winds, and the new temperature and
gravity calibrations by \citet{VGS96}. To derive the \Ha luminosity,
we sum up the emerging Lyman continuum photon emission
$N(\mathrm{H^0})$ for each star, and convert it to a luminosity in \Ha
using
\begin{equation}
   L(\mHa) \, \left[\mathrm{erg\,s}^{ -1 }\right] = 1.36\cdot10^{-12} \, N(\mathrm{H^0}) \left[\mathrm{s}^{ - 1 }\right].
   \label{eq:LymCont}
\end{equation}
We have updated our models using the emission line ratios for low
metallicities observed by \citet{ITL97} and \citet{IT98} for a large
sample of blue compact dwarfs (BCDs).

We present two types of one-zone models. The first represents a
quiescent galaxy, where the SFR decreases slowly from the formation
epoch with a timescale of 10\,Gyrs. The other model was already
discussed in detail by \citet{WDF+00} in their interpretation of Tidal
Dwarf Galaxy candidates, and can also be used to model blue compact
dwarfs \citep[BCDs, see][]{KFL95}. Here we put a starburst on top of
the stellar population of the undisturbed model.  It is assumed to
reach its maximum SFR of 20 M$_\odot\,$yr$^{-1}$ after 10\,Gyrs, and
we vary the burst timescale $\tau_\mathrm{B}$ from $10^5$ to
$10^8\,$yrs to model all the range of burst durations from small dwarf
galaxies to mergers of giant gas-rich galaxies. We alternatively a
bell shaped (Gaussian) burst or sharply rising and exponentially
decreasing starburst. We assume a minimum SFR after the starburst of
0.1\,M$_\odot\,$yr$^{-1}$.

%__________________________________________________________________

\section{Effects of Metallicity}\label{sec:metal}

We successfully reproduce the calibrations of HG86 and KTC94 using our
quiescent models using solar metallicity and upper mass limits of 120
and 85 M$_\odot$, respectively. The values we derive for the four
other metallicities are given in Tab.~\ref{tab:calib}.

\begin{table}[htbp]
  \begin{center}
    \caption{Calibrations for different metallicities $Z$.}
    \begin{tabular}{l r | c @{\hspace*{0.6cm}} c }
      \hspace*{+8pt}$Z$
            & $M_\mathrm{up}$
                  & \tcfac
                          & $\cfac_\mathrm{[\ion{O}{ii}]}$ \\
            & $\left[\mathrm{M}_\odot\right]$
                  & \multicolumn{2}{c}{$\left[10^{41}\,\frac{\mathrm{erg\,s}^{-1}}{\mathrm{M}_\odot\,\mathrm{yr}}\right]$} \\
      \hline
%     Z5..Z3     Z2        Z1    
%OII  3.010     1.791     1.303
%Ha   2.860     2.860     2.860
      0.001 &  85 & 4.221 & 1.923 \\
      0.001 & 120 & 4.702 & 2.142 \\
      0.004 &  85 & 3.105 & 1.944 \\
      0.004 & 120 & 3.315 & 2.076 \\
      0.008 &  85 & 2.073 & 2.182 \\
      0.008 & 120 & 2.548 & 2.682 \\
      0.020 &  85 & 1.229 & 1.293 \\
      0.020 & 120 & 1.408 & 1.482 \\
      0.040 &  85 & 0.902 & 0.949 \\
      0.040 & 120 & 0.997 & 1.049 \\
      \hline
      0.020 (S86)\hspace*{-8pt}
            & 120 & 0.917 & 0.965 \\[.5em]
      HG86  & 100 & 1.41  &  --- \\
      KTC94 & 100 & 1.26  &  --- \\
    \end{tabular}
    \label{tab:calib}
  \end{center}
\end{table}

%As the metallicity of spiral disks is usually around 1/2 $Z_\odot$ to $Z_\odot$ it does make sense that the lowest metallicity model is farthest away from the calibration of KTC94.
%dwarf irregulars ($\left<Z\right>_\mathrm{dIrr}=2-30\,\% Z_\odot$) or
While for solar metallicity models the agreement with HG86's and
KTC94's empirical calibrations is very good, it is also seen in
Tab.~\ref{tab:calib} that both towards lower and higher metallicities
the differences become significant. For subsolar metallicities as
e.g.~in BCDs \citep[$\left<Z\right>_\mathrm{BCD}= 0.002\pm0.001 = 1/10
Z_\odot$,][]{IT98} the difference in \tcfac is found to be a factor of
$\sim$3.5. This is a result of low metallicity stellar populations
being both more luminous and hotter than a solar metallicity stellar
population with the same IMF and mass limits.

Application of the empirical calibrations to estimate SFRs from \Ha
luminosities in low metallicity dwarf galaxies hence can yield SFRs
overestimated by a factor $\gtrsim$3 due to metallicity effects. We
also give in Tab.~\ref{tab:calib} the \tcfac-value obtained in our
solar metallicity model using the IMF of \citet{Sca86} instead of
Salpeter's. Due to the smaller number of high mass stars in case of a
Scalo-IMF, the same SFR produces an \Ha luminosity that is lower by
35\%.

All other hydrogen lines, as e.g.~L$_\alpha$ or Br$_\gamma$, which can
also be used to estimate SFRs, and their respective calibration factors
can easily be computed from our values for \Ha and Eq.~(\ref{eq:Ha}).

%To be able to use a SFR indicator for observations of sub-galactic fragments and Lyman break galaxies at higher redshift for non-solar metallicities, we also give the
At higher redshift, the [\ion{O}{ii}]3727 line is frequently used to
derive SFRs, as e.g.~for Lyman Break galaxies. This line involves a
direct metallicity dependence in addition to the one inherent in the
hydrogen lines which is due to differences in temperature and
luminosity of stellar populations at various metallicities.  Since
high redshift star-forming objects and, in particular, sub-galactic
fragments both have low metallicities and possibly strongly
fluctuating SFRs, care is needed to derive SFRs from
$L(\mathrm{[\ion{O}{ii}]})$. We present the values of the
$\cfac_\mathrm{[\ion{O}{ii}]}$-factor in Col.~4 of
Tab.~\ref{tab:calib} for the [\ion{O}{ii}]3727 line, where the
according linear relation with SFR is
\begin{equation}
   L\left(\mathrm{[\ion{O}{ii}]}\right) \, \left[\mathrm{erg\,s}^{-1}\right] = \cfac_\mathrm{[\ion{O}{ii}]} \cdot \mathrm{SFR}\,\left[\mathrm{M}_\odot\,\mathrm{yr}^{-1}\right].
   \label{eq:OII}
\end{equation}

%__________________________________________________________________

\section{Short starbursts}\label{sec:shortb}

\begin{figure}
  \centering
  \epsfig{file=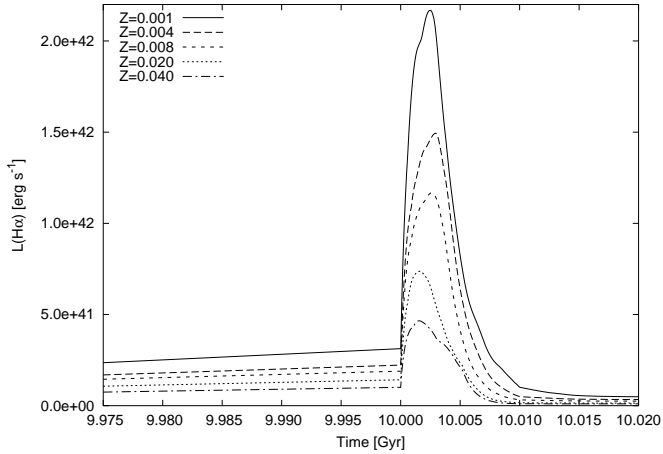,width=\linewidth}
  \caption{\LHa over time for instantaneous starbursts with short 
    timescale of $\tau_\mathrm{B} = 10^6\,$yr.}
  \label{fig:inst}%
\end{figure}

In Fig.~\ref{fig:inst} we show the time evolution of \LHa for rapidly
rising (instantaneous) starbursts with short timescales
($\tau_\mathrm{B} = 10^6\,$yr) for five metallicities. All models have
their maximum SFR at a time of 10.0\,Gyrs. We note two effects: For
lower metallicity $Z$ the maximum \Ha luminosity is higher than for
high metallicity, e.g.~by a factor of 4.6 in case of $Z=0.001$ as
compared with $Z=0.040$. This is due to the higher temperatures of the
low metallicity stars, which more efficiently ionize the interstellar
medium.  It is also apparent that there is a delay of the maximum in
the \Ha luminosity with regard to the maximum of the SFR. The offset
between maximum SFR and maximum \LHa is higher for lower
metallicities, 2.9\,Myrs for $Z=0.004$ vs.~1.5\,Myrs for $Z=0.040$.
This again is a result of the higher temperatures and hence the
ionizing power of low metallicity stars. At low metallicity stars of
lower mass and longer main sequence lifetimes contribute to the Lyman
continuum emission. Since those take longer to reach their maximum
Lyman continuum emission rates during their supergiant phase the delay 
becomes slightly longer than in the solar metallicity case.
%which therefore only becomes maximal somewhat longer after the sudden increase of the SFR than in of solar metallicity.

\begin{figure}
  \centering
  \epsfig{file=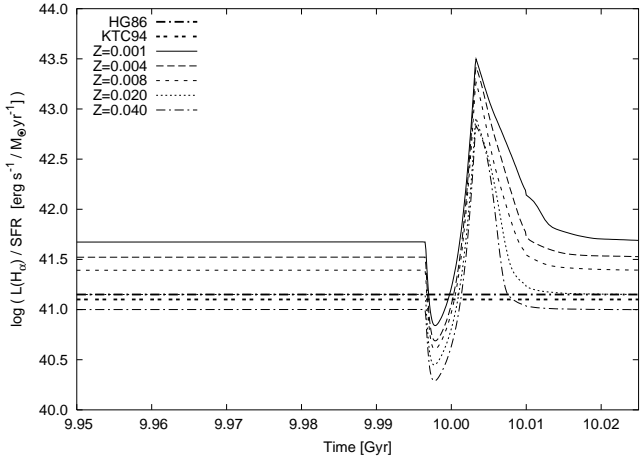,width=\linewidth}
  \caption{\tcfac = \LHa / SFR over time for Gaussian bursts with short 
    timescale of $\tau_\mathrm{B} = 10^6\,$yr.}
  \label{fig:time}%
\end{figure}

In Fig.~\ref{fig:time} we plot the ratio between \LHa and the SFR of
our models as a function of time for Gauss-shaped bursts with a short
timescale of $\tau_\mathrm{B} = 10^6\,$yr. Before and after the starburst
the empirical linear calibrations agree well with the ratio seen in our
solar metallicity model. With the onset of the burst the SFR rises faster
than \LHa due to the delay in the maximum Lyman continuum photon production
shown before in Fig.~\ref{fig:inst}. Therefore the \tcfac-factor first {\it
decreases}. After its minimum \tcfac~{\it increases}, because \LHa continues
to rise until the ionizing stars have reached their supergiant phase. When
the number of Lyman continuum photons and therefore \LHa become maximal,
the SFR has already decreased from its maximum value by a factor of $\sim$3.
After its maximum \LHa declines as the death rate of O stars is no longer
compensated by SF. In this phase, however, the SFR decreases even faster
than the \Ha flux, and \tcfac~{\it continues to increase}.  The sharp peak
visible near 10.003 Gyrs, after which \tcfac~{\it abruptly decreases}, is a
result of the constant minimum SFR of our models after the burst. If the SFR
after the burst would go to zero, \tcfac would diverge.  The minimum SFR,
which starts near 10.003 Gyrs, then acts to slowly {\it bring down}~\tcfac
to values in agreement with those in Tab.~\ref{tab:calib}. This time sequence
during the starburst and the interplay between \LHa and SFR can also be seen
in Fig.~\ref{fig:loop} below.

It is obvious from Fig.~\ref{fig:time} that for bursts with
$\tau_\mathrm{B} = 10^6\,$yr strong discrepancies from the calibrations
are seen for all metallicities, strongest at the lowest metal abundance,
amounting to a difference of nearly two orders of magnitude as compared
to the calibrations, even for solar metallicity models.

\begin{figure}
  \centering
  \epsfig{file=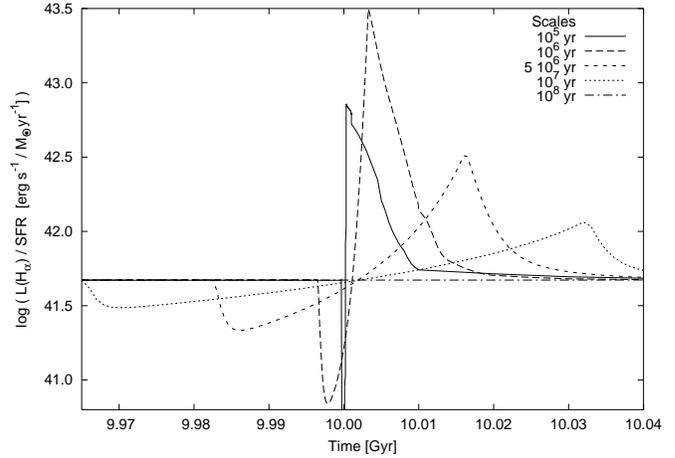,width=\linewidth}
  \caption{\tcfac = \LHa / SFR over time for Gaussian bursts 
    with different timescales and $Z = 0.001$.}
  \label{fig:scale}%
\end{figure}

In Fig.~\ref{fig:scale} we show the evolution of the \tcfac-factor
with time for five Gaussian starburst models with different burst
timescales for our lowest metallicity $Z=0.001$.  With increasing
burst duration, the maximum is shifted towards later times. It is
strongest for bursts with $\tau_\mathrm{B} = 10^6\,$yr (which are
shown in Fig.~\ref{fig:time}). This is a result of the convolution of
the Gauss-shaped increase of the burst-SFR with the delay in maximum
\Ha emission due to the most massive and most luminous (= giant)
stars. The decline after the sharply peaked maximum is an effect of
the onset of the minimum SFR as discussed above for Fig.~\ref{fig:time}.
The time evolution of \tcfac significantly depends on the burst
duration. For the longest burst with $\tau_\mathrm{B} = 10^8\,$yr the
slow change in the SFR of the burst causes the delay effect to have
negligible impact on \tcfac.

In conclusion we have shown that the SFR derived from \Ha during or
shortly after short bursts can be wrong by factors $\gg$10. Averaged
over the entire burst duration, however, values for \tcfac agree well
with those from Tab.~\ref{tab:calib} for different metallicities. 
This applies to statistical analyses of samples of \ion{H}{ii}
galaxies.

%__________________________________________________________________

\section{Effects of the IMF}\label{sec:imf}

\citet{KTC94} have already investigated different \tcfac-factors with
different IMFs with identical mass cutoffs and found differences of an
order of magnitude between IMFs they used. \citet{LH95} also presented
properties like number of O stars, $N(\mathrm{H}^0)$, and EW(\Ha)
etc.~for instantaneous and continuous SF models with different IMFs.
Here we want to show the effect of various IMFs on the time delay
discussed above.

\begin{figure}
  \centering \epsfig{file=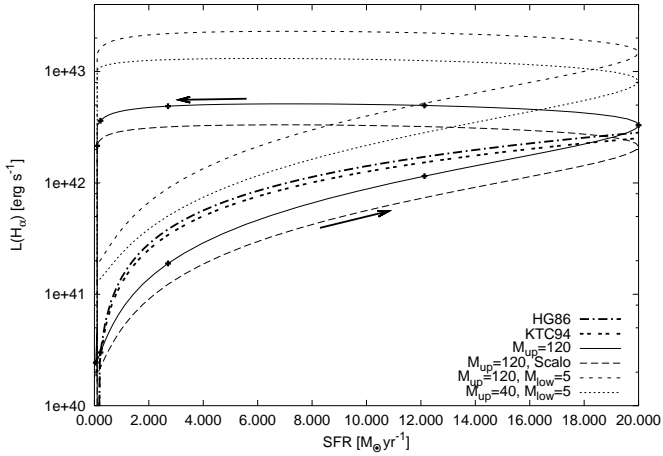,width=\linewidth}
  \caption{\LHa over SFR for a bursting dwarf galaxy with metallicity
    $Z=0.001$ and different IMF cuts.}
  \label{fig:loop}%
\end{figure}

Fig.~\ref{fig:loop} shows the luminosity \LHa plotted over the SFR of
the model galaxy, while it experiences its starburst with a timescale
of $\tau_B=10^6\,$yr. The calibrations of HG86 and KTC94 are plotted
for reference.  To indicate the evolution in time our ``standard
model'' with M$_\mathrm{up}=120$\,M$_\odot$ has additional dots
each $10^6\,$yrs during the burst; the arrows show the direction of
the loops in this diagram.

It is obvious that the calibrations do not represent the real SFR very
well.  \LHa at the peak SFR does fit very well with the calibrations
for both models with Salpeter and Scalo IMF and a high mass cut at
M$_\mathrm{up}=120$\,M$_\odot$. But generally our models show that for
an observed value of \LHa there is an ambiguity between two values of
the SFR. E.g.~for $L(\mHa) = 10^{42}\,$erg\,s$^{-1}$ the calibrations
give a SFR of $7 \dots 8\,$M$_\odot\,$yr$^{-1}$, while one has to
choose between SFRs of $\sim 0$ and $\sim 15\,$M$_\odot\,$yr$^{-1}$
from our model with Salpeter IMF, and one needs additional data to
determine the true current SFR.

The evolution of the models after the starburst shows that for low
SFRs (lower as one would expect from the calibrations) a galaxy could
show high \Ha luminosity for quite some time ($\sim 3\cdot10^6\,$yr).
For the two models with Salpeter IMF and lower mass cuts at
M$_\mathrm{low}=5$\,M$_\odot$ this effect is even more extreme. \LHa
at the peak SFR is already underestimated by the calibrations by
factors of 3 and 5, respectively, for high mass cutoffs of
M$_\mathrm{up}=40$ and $120$\,M$_\odot$. For a given \Ha-luminosity
the calibrations yield a SFR too high by at least one order of
magnitude when compared to the values from our models.

%__________________________________________________________________

\section{Conclusions}\label{sec:concl}

When observing small scale star-forming entities like dwarf galaxies
or sub-galactic fragments, where dynamical timescales and hence burst
durations may typically be of the order of $10^6\,$yrs or less, one
should be aware that the currently used calibrations on the basis of
\Ha luminosities may yield SFRs with large errors.  These calibrations
were determined from larger low-level star-forming systems (where they
work very well), and when applying them to star-bursting dwarf
galaxies, errors of factors 3 to 100 may affect this determination of
the SFR. The ratio of \Ha-luminosity to SFR depends on the metallicity
of the object, and the age and the duration of the starburst.

To more accurately determine the true current SFR of a (star-bursting)
dwarf galaxy additional information about the spectral energy
distribution (SED) is needed. Observations in {\it at least} three
optical/NIR filters or a spectrum with sufficient wavelength
coverage to determine the slope of the SED could be used in comparison
with models that include gaseous emission to eliminate ambiguities in
the relation of the \Ha luminosity to the SFR.

%__________________________________________________________________

\begin{acknowledgements}
  We thank our anonymous referee for a very prompt and helpful report.
  PMW is partially supported by Deutsche Forschungsgemeinschaft
  (DFG Grant FR 916/6-1).
\end{acknowledgements}

%__________________________________________________________________

\vspace*{-20pt}
\bibliography{../../PmW}
\bibliographystyle{apj}

\end{document}